\documentclass[12pt]{article}
\usepackage{latexsym}
\usepackage[dvips]{graphicx}
\usepackage{epsfig}
\def\be {\begin{equation}}
\def\beq {\begin{equation}}
\def\ee {\end{equation}}
\def\feq {\end{equation}}
\def\ba {\begin{eqnarray}}
\def\ea {\end{eqnarray}}

\def\bi {\begin{itemize}}
\def\ei {\end{itemize}}
\begin{document}
\def\bea{\begin{eqnarray}}
\def\eea{\end{eqnarray}}

\title{  BPS-like bound and thermodynamics of the charged BTZ black hole}
\author{ Mariano Cadoni\footnote{email: mariano.cadoni@ca.infn.it
}\, and  Cristina Monni \footnote{email: cristina.monni@ca.infn.it
}
\\
{\it \small Dipartimento di Fisica, Universit\`a di Cagliari and
INFN, Sezione di Cagliari }\\
{\small Cittadella Universitaria, 09042 Monserrato, Italy}}
\vfill

\maketitle

\maketitle

{\bf Abstract.}
The charged Ba\~nados-Teitelboim-Zanelli (BTZ) black hole is plagued 
by several  pathologies: 
$a$) Presence of divergent 
boundary terms in the action, hence of a  divergent black hole 
mass; $b$) Once a finite, renormalized, mass $M$ is defined 
black hole states exist for arbitrarily negative values of $M$; $c$) 
There is no upper bound on the charge $Q$.
We show that these  pathological features are an artifact of the 
renormalization procedure. They can be completely removed by  using  
an alternative  renormalization scheme leading to a different definition 
$M_{0}$
of the black hole mass,
which is the total energy inside the horizon. The new mass satisfies 
a BPS-like bound $M_{0}\ge \frac{\pi}{2}Q^{2}$ and the heat 
capacity  of the hole is positive. We also discuss the black hole
thermodynamics that arises when $M_{0}$ is interpreted as the 
internal energy of the system. We show, using three independent 
approaches (black hole thermodynamics, Einstein equations, 
Euclidean action formulation) that $M_{0}$ satisfies the first law if 
a term describing the mechanical work done by the electrostatic 
pressure is introduced.

\section{Introduction}

The discovery  of black hole solutions in three-dimensional (3D) anti 
de Sitter (AdS) spacetime by Ba\~nados, Teitelboim and Zanelli (BTZ)
\cite{Banados:1992wn,Banados:1992gq} (for a review see Ref.
\cite{Carlip:1995qv})  enhanced our understanding of 
black holes and  also played a  key role in recent developments in
gravity, gauge and string theory.
From the point of view of black hole physics,
the main lesson is that black holes can be formed by a 
singularity of the causal structure and not necessarily by 
a curvature singularity.  From the point of view of 
the AdS/CFT correspondence, the BTZ black hole is the simplest 
realization of a 3D bulk gravity configuration that can be described 
by dual thermal CFT states \cite{Brown:1986nw,Witten:2007kt}. One of the most striking 
successes of this correspondence is the exact computation of the 
Bekenstein-Hawking entropy of the BTZ black hole using the dual 
two-dimensional CFT \cite{Strominger:1997eq}.

It was  immediately realized that the BTZ solution allows also 
for charged generalizations, i.e charged black hole in 3D AdS 
spacetime  \cite{Banados:1992wn,Martinez:1999qi}. 
Differently from the uncharged case, these  black holes have a 
power-law curvature singularity and  share the causal structure with 
their higher-dimensional cousins (e.g. the charged Reissner-Nordstrom 
solution in 4D AdS space).

Naively, one could  expect that the analogy between 3D and 
higher-dimensional charged  black holes can be pushed forward to 
cover the main physical features of the black hole, such as the mass 
spectrum, thermodynamics etc. In particular, this analogy could be 
very useful for investigating in a simplified context peculiar 
features of the charged black holes, such as the existence of extremal 
black holes states of zero temperature and non-vanishing entropy. 

However,  this seems not to be the case, at least at first sight.
The charged BTZ black 
hole is plagued by several pathologies, which make it rather 
different from its higher-dimensional counterparts. The origin of 
these 
pathologies is well-understood; it can be traced back to the  
logarithmic behavior  of 
harmonic functions in two dimensions, which  
implies that the 
electrostatic potential of the charged BTZ black hole diverges 
asymptotically  as $\ln r$. The consequences are:
a) When we vary the action we get divergent 
boundary terms, i.e we have a divergent black hole 
mass; b) Using a suitable renormalization procedure, we can define a
finite  mass $M$ for the solution, but we find that 
black hole state exist for arbitrarily negative values of $M$; c) 
At  fixed $M$  there is no upper bound on the charge $Q$.
A further, recently discovered, manifestation of this problematic 
behavior, is the fact that the entropy function approach do not work 
for the extremal charged BTZ black hole \cite{Myung:2009sh}. 

These features make the charged BTZ black hole drastically different 
from the charged solutions in four and higher dimensions. In the 
latter case the black hole mass satisfies a BPS bound $M\ge a^{2} 
Q^{2}$, which guarantees that the mass is positive definite and that 
for a given mass the charge is bounded from above. The existence of 
this bound is usually a consequence of the supersymmetry of the 
extremal black hole.

Recently, an alternative renormalization procedure leading to a 
finite value $M_{0}$ for the  mass of the charged BTZ 
black hole, has been proposed 
\cite{Cadoni:2007ck,Cadoni:2008mw,Myung:2008kd}. Physically, $M_{0}$ is   
the total energy (gravitational 
and electromagnetic) inside the black hole outer horizon.
Moreover, the identification of $M_{0}$ with the conserved charge 
associated with time-translations is very natural from the point of 
view of the AdS/CFT correspondence \cite{Cadoni:2007ck}. It allows to 
reproduce, using the dual CFT,  the 
Bekenstein-Hawking entropy of the hole and to consider  the charged 
BTZ geometry as a bridge between two AdS$_{2}$ geometries, a near 
horizon one (AdS$_{2}\times S^{1}$) and an asymptotic one (linear 
dilaton  AdS$_{2}$) \cite{Cadoni:2007ck,Cadoni:2008mw}.

In view of these results, one is led to ask if the use of this 
alternative renormalization procedure for the mass, allows also to cure 
the pathologies of the charged  BTZ black hole.
 In this paper we investigate this issue.
We will show  that all the problematic features of the charged BTZ 
black hole can be removed if one uses  $M_{0}$ as black hole mass.
We will demonstrate that  $M_{0}$ satisfies 
a BPS-like bound $M_{0}\ge \frac{\pi}{2}Q^{2}$ and that for a black 
hole above extremality  the heat 
capacity   is positive and becomes zero in the extremal case.
We also discuss the formulation of  black hole
thermodynamics  when $M_{0}$ is interpreted as the 
internal energy of the thermodynamical system. We show, using three independent 
methods (black hole thermodynamics, Einstein equations, 
Euclidean action formulation) that $M_{0}$ satisfies the first law if 
a term describing the mechanical work done by the electrostatic 
pressure is introduced.

The structure of the paper is as follows. In Sect. 2 we review 
briefly the main features and pathologies of the charged BTZ black 
hole. In Sect. 3 we 
discuss in detail  the 
two renormalization schemes that can be used to get a finite black 
hole mass and show that the mass $M_{0}$ satisfies a BPS-like bound.
In Sect. 4 we discuss the thermodynamics of the charged BTZ black 
hole in the two cases, when the internal energy of the system is  
identified with either $M$ or $M_{0}$. In Sect. 5  we present our conclusions.

\section{The charged BTZ black hole}
The charged BTZ black hole 
solution is a U(1) generalization  of the uncharged
BTZ black hole \cite{Banados:1992wn}.
In this paper the solution with a non zero
electric charge $Q$ and zero angular momentum will be considered.

The action is 
\begin{equation}
 I = \int d^{3}x \sqrt{-g}\left(
\frac{R+2\Lambda}{2\pi}-\frac{1}{4}F^{\mu \nu }F_{\mu\nu}\right),
\label{azione}
\end{equation}
where $F_{\mu\nu}$ is Maxwell tensor, $\Lambda=1/l^{2}$ is the
cosmological 
constant ($l$ is the AdS length) and we are using units such that 
3D Newton constant $G$ is dimensionless,
$G=\frac{1}{8}$.

The solution for the electrically charged, non rotating case is given
by
\cite{Martinez:1999qi,Cadoni:2008mw}

\begin{equation}
ds^{2}=-f(r)dt^{2}+f^{-1}dr^{2}+r^{2}d\varphi^{2},
\label{solution}
\end{equation}
\begin{center}
  $0\leq r < \infty,    \hspace{5pt}   0\leq \varphi < 2\pi,
\hspace{5pt}$
 \end{center}
with metric function and Maxwell field
\begin{equation}
 f = \frac{r^{2}}{l^{2}}-M-\pi Q^{2}\ln \frac{r}{w}, \quad 
 F_{tr}=\frac{Q}{r},
\label{f} 
\end{equation} 
where $M,Q,w$ are integration constants. Although the solution 
depends only on two integration constants ($w$  can be absorbed in a 
redefinition of $M$), it is convenient to keep the dependence of the 
metric on the arbitrary length scale $w$ explicit.
 The above  solution  represents a 3D, asymptotically AdS,  black hole,  
 with a power-law singularity at $r=0$, where $R\sim \pi Q^{2}/r^{2}$ 
 and, generically, with an inner ($r_{-}$) and outer ($r_{+}$) horizon.
$M,Q$ could be naively seen as 
the black hole mass and charge,  respectively.
$M$ can be easily expressed as a function of the charge and of
the outer horizon radius,
\begin{equation}
 M(r_{+},Q,w) = \frac{r_{+}^{2}}{l^{2}}-
 \pi Q^{2}\ln\left( \frac{r_{+}}{w}\right).
\label{M}
\end{equation} 
Whereas the 
interpretation of $Q$ as the black hole electric charge is 
straightforward, the same is not true for $M$.
In fact by varying the action (\ref{azione}), one finds a surface
term which diverges in the
limit $r\rightarrow\infty$ \cite{Martinez:1999qi}:
\begin{equation}
 \left( -\delta M -\pi \delta Q^{2}\ln r\right) N(r),
\label{surfaceterm} 
\end{equation}
where $N$ is the lapse function.
The presence of the logarithmic  divergent boundary term 
makes the black hole mass a poorly 
defined concept.

\begin{figure}
   \center
  \includegraphics[width=300pt]{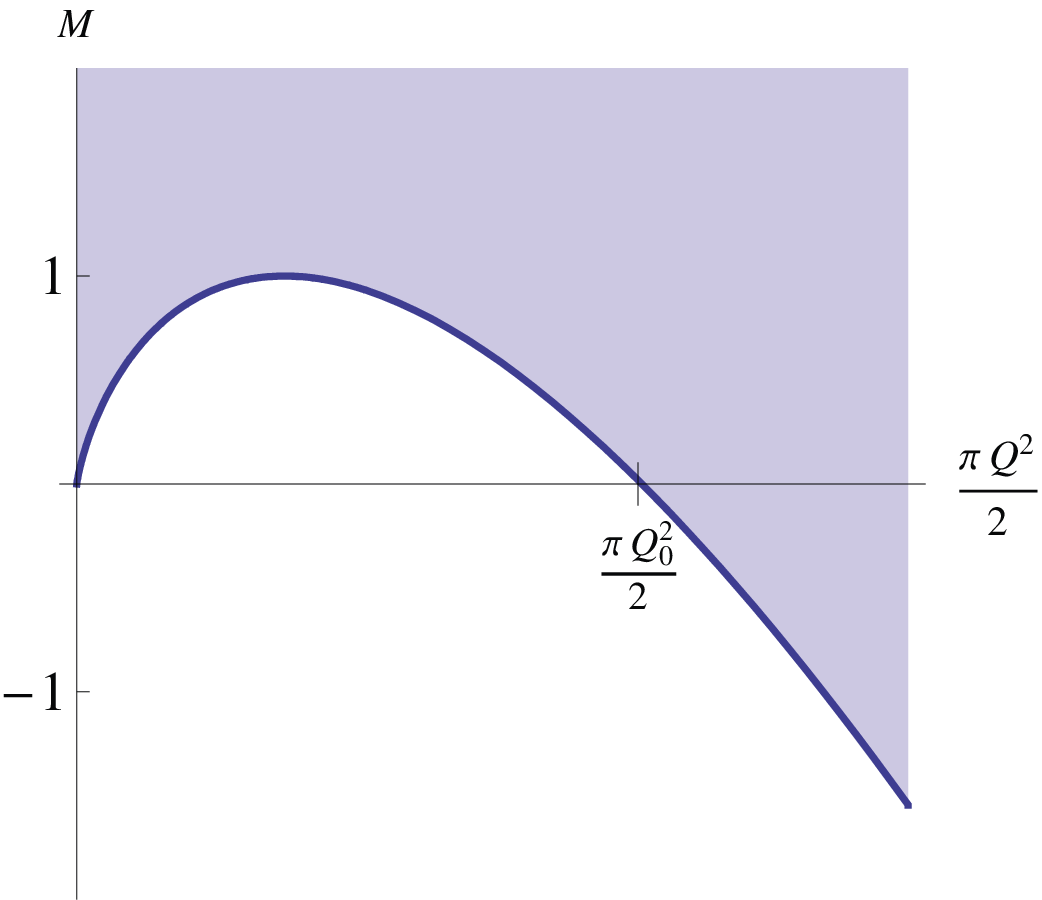}\\
 \caption{ Region in the  $M-Q$ phase space  where the
black hole exists. The region of existence 
is the shaded region. Extremal black holes are in the boundary line
between the shaded and the unshaded regions.}
 \label{MvsQ}
\end{figure}

A second unpleasant feature emerges when one imposes a {\it cosmic 
censorship} condition, i.e the absence of naked singularities.
The requirement that   the singularity at $r=0$ is shielded by an
event 
horizon is equivalent 
to requiring that the metric function $f(r)$ 
evaluated  on its minimum value, is equal or less
than zero (corresponding respectively to one or two horizons).
Introducing a function $\Delta(M,Q)$  
the condition for the existence of the horizon(s) can 
be cast in the form

\begin{equation}
\Delta(M,Q) = f\left( r=r_{min}=\sqrt{\frac{\pi}{2}}Ql\right) =- M +
\frac{\pi Q^{2}}{2}\left( 1-\ln\frac{\pi Q^{2}}{2}\right)\leq 0,
\label{delta}
\end{equation}
Eq. (\ref{delta}) is not a BPS-like bound, it can be satisfied by 
negative values of the mass $M$. 
This can be immediately seen considering 
the $M - Q$  phase diagram shown in Fig (\ref{MvsQ}). 

For $Q$ above a critical value $Q_{0}$ there are  black hole
solutions with 
arbitrarily negative values of $M$. 
The presence of black hole states with mass values unbounded from
below 
makes the system intrinsically unstable and the definition of 
thermodynamical ensembles problematic. 
At this point one can wonder if the presence of black hole states 
with arbitrary negative mass is a physical feature of the system or 
an artifact due to the divergence of the boundary term. 
This question can be answered only after the issue of mass 
renormalization is discussed in detail.

\section{Mass renormalization} 
The problem of the presence of a divergent boundary term can be 
handled  with a renormalization procedure.
This procedure can be implemented in  systematic way 
 by enclosing our system in a circle of radius 
$r_{0}$. This allows to rewrite the metric function  (\ref{f}) in the form
\begin{equation}\label{mf}
 f(r) = -M_{0}(r_{0}, w) + \frac{r^{2}}{l^{2}}-\pi Q^{2}\ln\left(
\frac{r}{r_{0}}\right) 
\end{equation} 
and to  define a regularized mass:
\begin{equation}\label{M0}
 M_{0}(r_{0}, w) = M + \pi Q^{2}\ln\left( \frac{r_{0}}{w}\right).
\end{equation} 
$w$ is now considered   a running scale and $M_{0}(r_{0})$ is the 
total energy (gravitational and electromagnetic) inside a circle of 
radius $r_{0}$; in the limit $r\rightarrow \infty$ 
one takes also $r_{0}\rightarrow \infty$,
keeping the ratio $r/r_{0}=1$   \cite{Cadoni:2007ck,Cadoni:2008mw}.
One is left with two possible options: i) $M$ is held fixed and the 
space-time metric is 
scale-dependent; ii) The metric is $w$-invariant and $M$ runs with
$w$.

Option i) is the renormalization procedure proposed in
\cite{Martinez:1999qi}.  
Eq. (\ref{M0}) is used to identify $M$ as the total mass of the 
solution, an infinite constant is absorbed in $M_{0}$ and the 
logarithmic divergent term in Eq. (\ref{surfaceterm}) is removed.
Apart from the drawback of having the metric (hence the 
horizons position and the  black hole entropy) running with $w$, this 
procedure  does not solve the stability problem, hence  does 
not allow for a consistent interpretation of the charged BTZ black 
hole as a thermodynamical system.

In this paper we use the $w$-invariant renormalization prescription
ii) 
first  proposed in Ref. \cite{Cadoni:2007ck}. To keep the 
metric  function (\ref{mf}) unchanged as $w$ runs, we held  
$M_{0}(r_{0},w)$ fixed, whereas $M$
changes with $w$:
$w\rightarrow \lambda w, \quad M \rightarrow M+ \pi Q^{2}\ln \lambda$.
 The boundary term (\ref{surfaceterm}) becomes  now
\begin{equation}
 \left( -\delta M_{0}-\pi \delta Q^{2} \ln\frac{r}{r_{0}}\right) N(r),
\end{equation} 
and in the limit $r,r_{0}\to \infty$  the divergent part is removed 

As a consequence of the $w$-invariance of $f$ and $M_{0}$, we can
arbitrarily
choose $w$ and $M_{0}$. Following Ref. \cite{Cadoni:2008mw} we choose 
to fix them in 
terms of the AdS length  ($w =
l$) and of  the horizon position ($r_{0}=r_{+}$).

This renormalization procedure allows us to  associate in a 
consistent way to every charged BTZ  black hole solution 
(\ref{solution}), (\ref{f})  a 
finite mass given by
\begin{equation}
 M_{0}(r_{+}) = M + \pi Q^{2}\ln\left( \frac{r_{+}}{l}\right).
\label{M_0}
\end{equation} 
The metric function becomes
\begin{equation}
 f(r, M_{0}) = -M_{0} +\frac{r^{2}}{l^{2}}-\pi
Q^{2}\ln\frac{r}{r_{+}}.
\label{p8}
\end{equation} 
It is important to stress that setting $r_{0}=r_{+}$ we are 
implicitly assuming that at least one horizon is always present, i.e. 
the validity of the cosmic censorship conjecture.

We can identify the $w$-invariant mass $M_{0}(r_{+})$ as
 the
conserved charge associated with time-translation invariance, instead
of the mass $M$.
The renormalization prescription ii) has further nice features.
The renormalized  mass $M_{0}$ depends only on the 
horizon position, is always positive-definite and shares with the
uncharged BTZ black hole the mass/horizon-position  dependence: 
\begin{equation}\label{e1}
 M_{0}(r_{+}) = \frac{r_{+}^{2}}{l^{2}}.
\end{equation} 
Moreover, the identification of $M_{0}$ with the conserved charge
associated 
with time-translation allows to reproduce exactly the 
Bekenstein-Hawking entropy of the charged BTZ black hole using  a 
Cardy formula for the 2D dual CFT \cite{Cadoni:2007ck}.

\subsection{BPS-like bound}

Let us now show that the use of $M_{0}$ as the physical mass of the 
system, allows, at least in principle, to solve the instability 
problem.
The new mass spectrum can be found using Eq. (\ref{e1}) and the 
trivial relation $r_{+}\ge r_{min}= \sqrt{\frac{\pi}{2}} Ql$.  We have
\begin{equation}
M_{0}\ge \frac{\pi}{2} Q^{2}.
\label{BPS_bound}
\end{equation}
Because by writing Eqs. (\ref{M_0}) and (\ref{p8}) we have implicitly 
assumed that at least one horizon is present, we expect the inequality 
(\ref{delta})   to be identically satisfied.
In fact, expressing
$\Delta$ of  Eq. (\ref{delta}) as a function of $M_{0}$,
\begin{equation}
  \Delta(M_{0},Q) = -M_{0}+\frac{\pi Q^{2}}{2}\left(
1-\ln\frac{\pi Q^{2}}{2M_{0}}\right) ,
\label{Delta1}
\end{equation} 
and setting $\alpha = 2M_{0}/\pi Q^{2}$ the condition for the existence 
of the horizons, $\Delta\leq 0$, takes   the form $\alpha -1 \ge \ln 
\alpha$, which is always  true.

Eq. (\ref{BPS_bound}) represents a BPS-like bound for the black hole 
mass. It takes a form similar to the bound satisfied by charged black
holes 
in higher dimensions.
The bound is saturated  when in Eq. 
(\ref{BPS_bound}) 
the equality holds. In this case we have an extremal black hole, 
which is a state of mass $M_{0}= (\pi/2) Q^{2}$, zero temperature and 
nonvanishing entropy $S=2\pi \sqrt{2\pi}\, lQ$ (see Eq. (\ref{TSP}) below).
Again, these are features shared by higher dimensional charged black 
holes. 
The quadratic form of the $M_{0}-Q$ phase diagram, which results from 
Eq. (\ref{BPS_bound}), eliminates the 
negative, unbounded from below, tail  present in Fig. \ref{MvsQ}
and sets an upper bound on the  black hole charge Q.

This result implies  that the presence of black hole states with 
arbitrary negative mass is a consequence of identifying the energy of 
the system with the mass $M$. 
Moreover, it gives a strong hint that 
the $M_{0}=\pi Q^{2}/2$ extremal black 
hole could be a stable configuration.
Obviously, in the context of our discussion
stability  is just a consequence of  the validity
of the cosmic censorship conjecture. 
A formal   proof of the stability of the extremal  configuration 
would require  a detailed analysis of the perturbation spectrum 
around the extremal black hole solution.  Alternatively, stability 
can be proved by showing that the extremal background is 
supersymmetric \cite{Kallosh:1992ii},
i.e. it allows for  the existence of Killing spinors.
A detailed analysis of the stability of the extremal black hole is 
outside the aim of this paper. In the next sections we will show 
that using $M_{0}$ as internal energy of the system allows for a consistent 
formulation of the
thermodynamics of the charged BTZ black hole.   

\section{The first law of thermodynamics for the charged BTZ black hole}
\label{sec:3}
It is well known that the laws of black hole mechanics  mimic 
the laws of thermodynamics. Formally, a black hole can be considered
a 
thermodynamical system. 
Obviously, the thermodynamical behavior of our charged BTZ black
hole 
will depend on the identification of the black hole parameters in 
terms of thermodynamical variables. For the temperature $T$, the 
entropy $S$, the electric potential $\Phi$ (thought of as chemical 
potential) there is no ambiguity. $T,S,\Phi$ are given as usual in
terms of,
respectively,
surface gravity, horizon area and time component of the vector 
potential,
\begin{eqnarray}
 T &=& \frac{1}{4\pi}\left( \frac{2r_{+}}{l^{2}}- \frac{\pi
Q^{2}}{r_{+}}\right),\nonumber\\ S &=& 4\pi r_{+}= 
4 \pi l\sqrt{ \pi Q^{2}\ln\frac{r_{+}}{l}+M}, \nonumber\\\Phi
&=&A_{0}(r_{+}) = -2\pi Q \ln\frac{r_{+}}{l}.
\label{TSP}
\end{eqnarray} 
On the other hand, for the internal  energy $E$ of the thermodynamical
system we have 
two possible choices: we can identify $E$ either with $M$ or with 
$M_{0}$.
We will show now that both choices lead, at least at the formal 
level,  to a consistent thermodynamical formulation.

For $E=M$ the internal energy of the system is the total 
energy (gravitational and electrostatic) of the black hole and we 
expect the first principle to take the usual form. 
In fact differentiating $M(r_{+},Q)$ in Eq. (\ref{M}) and making use
of Eqs. 
(\ref{TSP}) one easily obtains the first principle in the form
\begin{equation}
 dM = TdS + \Phi dQ.
\label{dM}
\end{equation} 

The exact form $M(S,Q)$ can be easily determined, we have
\begin{equation}
 M(S,Q) = \frac{S^{2}}{16 \pi^{2}l^{2}} -\pi Q^{2}\ln 
 \left(\frac{S}{4\pi l}\right).
\label{mexact}
\end{equation} 

Conversely, when $E=M_{0}$ the internal energy is identified with the 
energy of the black hole inside the radius $r_{+}$. In this  
case we  expect that the presence of radial pressure gradients will 
give rise to additional terms in Eq. (\ref{dM}).
The new form of the first principle can be obtained differentiating 
$M_{0}$ given in equation (\ref{M_0}) and 
using Eq. (\ref{dM}). One has,
\begin{equation}
  dM_{0} = TdS  +\Phi dQ +dK,
\label{dM0}
\end{equation} 
where $K= \pi Q^{2}\ln\left( \frac{r_{+}}{l}\right)$ is minus the 
electrostatic energy outside the horizon.
The variation of $K$ cannot   change the black hole 
entropy but represents  mechanical work done by electrostatic
pressure.
We can compute $dK$  keeping  constant the electrostatic 
potential $\Phi$, this allows to express charge variation in terms 
of displacement of the horizon,
\begin{equation}\label{df}
2 \pi \ln\frac{r_{+}}{l}dQ= - \frac{2 \pi Q}{r_{+}}dr_{+}. 
\end{equation}
Using the previous equation one finds
\begin{equation}\label{dM1}
  dM_{0} = TdS  +\Phi dQ -\frac{\pi Q^{2}}{r_{+}}dr_{+}.
\end{equation} 
The last term in the previous equation is the work done by the radial 
pressure 
\begin{equation}
 P_{r}= T_{r}^{r}
\label{rp}
\end{equation} 
generated by the electrostatic field
($T_{\mu\nu}$ is the stress-energy tensor for the Maxwell field).
Explicit computation of the $T_{r}^{r}$ gives,
\begin{equation}
 P_{r}(r_{+}) = -\frac{Q^{2}}{2r_{+}^{2}}.
\label{P}
\end{equation} 
Using  Eq. (\ref{P}) in (\ref{dM1}) one obtains the first principle in 
the final form \footnote{A first principle of this form for the charged BTZ black 
hole  has been also derived in Ref. \cite{Akbar:2007zz,Akbar:2007qg}. 
However, in those papers  a 
different definition for the  
internal energy $E$ is used. $E$ is not identified with the mass 
$M_{0}$,  but is given by  our Eq. 
(\ref{M_0}) with opposite sign of the second term in the r.h.s.} 
\begin{equation}
 dM_{0} =  TdS +\Phi dQ +P_{r}d\mathcal A,
\label{1law}
\end{equation} 
where $\mathcal A=\pi r_{+}^{2}$ is the area inside the radius
$r_{+}$.
Notice that when $d \mathcal A>0$ the  mechanical work   $P_{r}d \mathcal A$ in 
Eq. (\ref{1law}), is negative, i.e it is done by the thermodynamical 
system.  The
pressure $P_{r}(r_{+})$ goes to zero when
$r_{+}\rightarrow\infty$, in this situation  $M=M_{0}$
and the two thermodynamical descriptions are equivalent. 

The internal energy $M_{0}$ appearing in the first principle 
(\ref{1law}) appears   to be a function of three independent 
extensive thermodynamical parameters $S,Q,\mathcal A$; a simple 
calculation gives 
\begin{equation}
 M_{0}(S,Q,\mathcal A) = \frac{S^{2}}{16 \pi^{2}l^{2}} -\pi Q^{2}\ln 
 \left(\frac{S}{4\pi l}\right)+ \frac{\pi Q^{2}}{2} \ln
 \left(\frac{\mathcal 
 A}{\pi l^{2}}\right).
\label{m0exact}
\end{equation} 

However, there are only two independent parameters because of the 
presence of a constraint. This  constraint takes a different form 
for thermodynamical transformations at constant $\Phi$ or constant $Q$.
In the first case  the 
constraint takes the form (\ref{df}), which can be also written as
\begin{equation}
\label{c1}
\Phi dQ=-2 P_{r} d \mathcal A.
\end{equation} 
Conversely, keeping the charge $Q$ constant the constraint  takes the form
\begin{equation}
\label{c2}
Qd\Phi =-2  {\mathcal A} d P_{r} .
\end{equation}

It is also interesting to compute the thermal capacity of the black 
hole at constant 
charge  as a function of $M_{0}$. We have
\begin{equation}
 C= T\frac{\partial S}{\partial T}|_{Q}=4\pi
l\sqrt{M_{0}}\frac{2M_{0}-\pi Q^{2}}{2M_{0}+\pi Q^{2}}.
\label{cap_term}
\end{equation} 
The heat capacity is always positive when the black hole is above 
extremality, 
$M_{0}\ge \pi Q^{2}/2$, and becomes zero in the extremal case.

\subsection{Derivation of the  first law from Einstein's equations}
Black hole thermodynamics can be derived 
from the laws of black hole mechanics, i.e it is codified in Einstein 
equations.  The first principle of thermodynamics for the charged BTZ 
black hole (\ref{1law})
can be derived from the $rr$ component of Einstein's equation 
\cite{Padmanabhan:2003gd,Paranjape:2006ca},
\begin{equation}
 G^{r}_{r}-\Lambda g_{r}^{r} = \pi T_{r}^{r},
\label{eq.einstein}
\end{equation} 
where $G_{\mu\nu}=R_{\mu\nu}-\frac{1}{2}g_{\mu\nu}R$ is the Einstein tensor. 
Evaluating the equations (\ref{eq.einstein}) on the horizon and 
using Eq. (\ref{rp}) one has

\begin{equation}
 \frac{f'(r_{+})}{2r_{+}}-
\frac{1}{l^{2}} = \pi P_{r}.
\label{eqe}
\end{equation} 
Multiplying both terms of this equation by $d(r_{+}^{2})$ and
using 
Eqs. (\ref{TSP}) and (\ref{e1}) we obtain
\begin{equation}
 dM_{0} = TdS - P_{r}d\mathcal A.
\label{dM_0.1}
\end{equation} 
Using Eq. (\ref{c1}) 
we easily find that   equation (\ref{dM_0.1}) is equivalent to the first law
(\ref{1law}).

\subsection{Euclidean action formulation}

In this section we will show that the thermodynamics of the charged 
BTZ black hole described in the previous sections can be also derived 
using the Euclidean action formalism.
In the Euclidean action approach to black hole thermodynamics, one can get
Gibbs  free energy through analytic continuation of
the action, with suitable boundary terms, in Euclidean space.

To compute the Euclidean action $I_{E}$ we will follow 
the method of  Ba\~{n}ados, Teitelboim e Zanelli
\cite{Banados:1992wn}, which  use the Hamiltonian version of the action
(\ref{azione}). The bulk contribution  is
equal to zero  and the Euclidean action is
completely given by three surface terms.
The first surface term  must be added at infinity  and 
is  given by the mass of the solution times   the
periodicity  of  of Killing time $\beta=1/T$. 
The other  two surface terms  make sure that the variational 
derivative  of the action vanishes on the horizon.

If we use the regularization scheme i) of section 3 the boundary term 
at  infinity  is given 
by $M$ and all together  one has, for the Euclidean action:
\begin{equation}
 I_{E} = \beta M-4\pi r_{+}-\beta A_{0}(r_{+})Q.
\label{IE}
\end{equation} 
Gibbs free energy, $G(T,\Phi)$ describing the system in the grand canonical
ensemble, is given by $G= T I_{E}$ and using Eq. (\ref{IE}) it turns out 
to be, as expected,
the Legendre transform of $M$ with respect to $S$ and $Q$:
\begin{equation}\label{G}
 G(T,\Phi) = M -TS-\Phi Q.
\end{equation} 
The description of the thermodynamical system  trough $G(T,\Phi)$ 
corresponds to the choice of $M$ as the internal energy of the system.
One can easily reproduce the entropy $S$ and the charge $Q$ as
$S = -(\partial G)/(\partial T)$ and $Q = -(\partial G)/(\partial \Phi)$.

If we use instead the renormalization scheme ii) of section 
3, the Euclidean action (\ref {IE}), hence Gibbs free energy does not change. 
The mass of the solution is now  $M_{0}$ but the boundary term at 
infinity is still  given by
by $M$, the logarithmic term in Eq. (\ref{M_0})  being an  horizon
contribution. 
From  Eq. (\ref{M_0}) it follows that $M_{0}$ can be written  as
\begin{equation}\label{e3}
 M_{0}=M-\frac{1}{2} Q\Phi(r_{+}).
 \end{equation} 
Thus, the term needed to cancel the 
variational derivatives of the action on the horizon  
is now  given by $-4\pi r_{+}-(1/2)\beta Q\Phi$. All these contribution 
sum up to the same result given in  Eq. (\ref{IE}).

Corresponding to the choice of $M_{0}$ as the internal energy of the 
system, 
Gibbs free energy can be now expressed as a function of $T,\Phi, 
P_{r}$. This can be done by first  
making use of Eq. (\ref{e3}) to write $G$ in  equation (\ref{G}) in 
terms of $M_{0}$: $G=M_{0}-TS-\Phi Q +\frac{1}{2}\Phi Q$, then 
differentiating and using   
the first law (\ref{1law}) and the constraints (\ref{c1}), (\ref{c2}). 
One  obtains  
\begin{equation}
 dG(T,\Phi,P)=-SdT- Qd\Phi  - \mathcal AdP_{r}.
\end{equation} 

\section{Conclusions}
In this paper we have shown that the problematic features of the charged 
BTZ black hole  are an artifact of the usual  
renormalization procedure for the divergent bare mass of the hole.
An alternative renormalization scheme leads to a different definition 
of black hole mass, which physically is the total energy inside the 
horizon. When described in terms of this mass $M_{0}$, the charged BTZ black 
hole behaves like the 4D Reissner-Nordstrom black hole. It satisfies 
a BPS-like bound that guarantees both  positivity of the mass and an
upper bound for the charge of the hole. The extremal 
black hole is a state of zero temperature and nonvanishing entropy. 
The thermal capacity of the hole is always positive and becomes zero 
for the extremal black hole.

We have also shown, using three different approaches, that the charged BTZ 
black hole allows for a consistent thermodynamical description  
when $M_{0}$ is interpreted as the internal energy of the system. The 
only change with respect to usual black hole thermodynamics 
is the appearance in the first law of a term describing the 
mechanical work done by the electrostatic pressure.

These results improve our understanding  of charged black hole solution   
and could be also very useful in the AdS/CFT correspondence  context. 
Similarly to 
the higher-dimensional cases the BTZ black hole is a bridge 
between  a near-horizon AdS$_{2}$
 and  an asymptotic AdS$_{3}$  geometry 
\cite{Cadoni:2008mw}. This feature could be very useful for understanding the 
nature of AdS$_{2}$ quantum gravity and in particular the microscopic 
entropy of extremal charged black holes 
\cite{Cadoni:2008pr, Hartman:2008dq,Sen:2008yk,Gupta:2008ki,Castro:2008ms,Hotta:2008xt,
Myung:2009sk,Hotta:2009bm}.

In this paper we have not addressed at a full  level the issue 
of the stability of the extremal charged BTZ black hole.
The presence of the BPS-like bound (\ref{BPS_bound}), the 
vanishing of the temperature and of the thermal capacity in the extremal 
configuration strongly indicates that the extremal charged BTZ black 
hole is stable. However, a proof of this statement will require 
detailed analysis of the perturbation spectrum or, alternatively, the 
demonstration that this extremal state is a true supersymmetric BPS 
state.


\begin{thebibliography}{100}

\bibitem{Banados:1992wn}
  M.~Banados, C.~Teitelboim and J.~Zanelli,
  Phys.\ Rev.\ Lett.\  {\bf 69} (1992) 1849
  [arXiv:hep-th/9204099].
\bibitem{Banados:1992gq}
  M.~Banados, M.~Henneaux, C.~Teitelboim and J.~Zanelli,
  Phys.\ Rev.\  D {\bf 48} (1993) 1506
  [arXiv:gr-qc/9302012].

\bibitem{Carlip:1995qv}
  S.~Carlip,
  Class.\ Quant.\ Grav.\  {\bf 12} (1995) 2853
  [arXiv:gr-qc/9506079].


    
    
\bibitem{Brown:1986nw}
  J.~D.~Brown and M.~Henneaux,
  Commun.\ Math.\ Phys.\  {\bf 104}, 207 (1986).

\bibitem{Witten:2007kt}
  E.~Witten,
  arXiv:0706.3359 [hep-th].

\bibitem{Strominger:1997eq}
  A.~Strominger,
  JHEP {\bf 9802} (1998) 009
  [arXiv:hep-th/9712251].

\bibitem{Martinez:1999qi}
  C.~Martinez, C.~Teitelboim and J.~Zanelli,
  Phys.\ Rev.\  D {\bf 61} (2000) 104013
  [arXiv:hep-th/9912259].
  
\bibitem{Myung:2009sh}
  Y.~S.~Myung, Y.~W.~Kim and Y.~J.~Park,
  arXiv:0903.2109 [hep-th].

\bibitem{Cadoni:2007ck}
  M.~Cadoni, M.~Melis and M.~R.~Setare,
  Class.\ Quant.\ Grav.\  {\bf 25} (2008) 195022
  [arXiv:0710.3009 [hep-th]].


\bibitem{Cadoni:2008mw}
  M.~Cadoni and M.~R.~Setare,
  JHEP {\bf 0807} (2008) 131
  [arXiv:0806.2754 [hep-th]].

\bibitem{Myung:2008kd}
  Y.~S.~Myung, Y.~W.~Kim and Y.~J.~Park,
  Phys.\ Rev.\  D {\bf 78} (2008) 044020
  [arXiv:0804.0301 [gr-qc]].

\bibitem{Kallosh:1992ii}
  R.~Kallosh, A.~D.~Linde, T.~Ortin, A.~W.~Peet and A.~Van Proeyen,
  Phys.\ Rev.\  D {\bf 46} (1992) 5278
  [arXiv:hep-th/9205027].

\bibitem{Akbar:2007zz}
  M.~Akbar and A.~A.~Siddiqui,
  Phys.\ Lett.\  B {\bf 656} (2007) 217.

\bibitem{Akbar:2007qg}
  M.~Akbar,
  Chin.\ Phys.\ Lett.\  {\bf 24} (2007) 1158
  [arXiv:hep-th/0702029].
 
\bibitem{Padmanabhan:2003gd}
  T.~Padmanabhan,
  Phys.\ Rept.\  {\bf 406} (2005) 49
  [arXiv:gr-qc/0311036].

\bibitem{Paranjape:2006ca}
  A.~Paranjape, S.~Sarkar and T.~Padmanabhan,
  Phys.\ Rev.\  D {\bf 74} (2006) 104015
  [arXiv:hep-th/0607240].

  
\bibitem{Cadoni:2008pr}
  M.~Cadoni, M.~Melis and P.~Pani,
  PoS {\bf BHs,GR and Strings} (2009) 032
  [arXiv:0812.3362 [hep-th]].
\bibitem{Hartman:2008dq}
  T.~Hartman and A.~Strominger,
  JHEP {\bf 0904} (2009) 026
  [arXiv:0803.3621 [hep-th]].

  \bibitem{Sen:2008yk}
  A.~Sen,
  JHEP {\bf 0811} (2008) 075
  [arXiv:0805.0095 [hep-th]].

\bibitem{Gupta:2008ki}
  R.~K.~Gupta and A.~Sen,
  JHEP {\bf 0904} (2009) 034
  [arXiv:0806.0053 [hep-th]].



\bibitem{Castro:2008ms}
  A.~Castro, D.~Grumiller, F.~Larsen and R.~McNees,
  JHEP {\bf 0811} (2008) 052
  [arXiv:0809.4264 [hep-th]].

\bibitem{Hotta:2008xt}
  K.~Hotta, Y.~Hyakutake, T.~Kubota, T.~Nishinaka and H.~Tanida,
  JHEP {\bf 0901} (2009) 010
  [arXiv:0811.0910 [hep-th]].

 
\bibitem{Myung:2009sk}
  Y.~S.~Myung, Y.~W.~Kim and Y.~J.~Park,
  arXiv:0901.2141 [hep-th].

  \bibitem{Hotta:2009bm}
  K.~Hotta,
  arXiv:0902.3529 [hep-th].




\end{thebibliography}
\end{document}